\documentclass[aps,prl,groupedaddres,amssymb,twocolumn,superscriptaddress]{revtex4-1}
\usepackage{amsmath}
\usepackage{graphicx}
\usepackage{wasysym,colortbl}

\begin{document}


\title{An improved measurement of the Lense-Thirring precession on the orbits of  laser-ranged satellites with an {accuracy} approaching the 1\% level}


\author{David M. Lucchesi}
\email[]{david.lucchesi@inaf.it}
\affiliation{Istituto di Astrofisica e Planetologia Spaziali (IAPS) - Istituto Nazionale di Astrofisica (INAF)\\Via del Fosso del Cavaliere 100, 00133 Roma, Italy}
\affiliation{Istituto Nazionale di Fisica Nucleare (INFN), Sezione di Roma Tor Vergata\\Via della Ricerca Scientifica 1, 00133 Roma, Italy}
\affiliation{Istituto di Scienza e Tecnologie dell'Informazione (ISTI) - Consiglio Nazionale delle Ricerche (CNR)\\ Via Moruzzi 1, 56124 Pisa, Italy}
\author{Massimo Visco}
\affiliation{Istituto di Astrofisica e Planetologia Spaziali (IAPS) - Istituto Nazionale di Astrofisica (INAF)\\Via del Fosso del Cavaliere 100, 00133 Roma, Italy}
\affiliation{Istituto Nazionale di Fisica Nucleare (INFN), Sezione di Roma Tor Vergata\\Via della Ricerca Scientifica 1, 00133 Roma, Italy}
\author{Roberto Peron}
\affiliation{Istituto di Astrofisica e Planetologia Spaziali (IAPS) - Istituto Nazionale di Astrofisica (INAF)\\Via del Fosso del Cavaliere 100, 00133 Roma, Italy}
\affiliation{Istituto Nazionale di Fisica Nucleare (INFN), Sezione di Roma Tor Vergata\\Via della Ricerca Scientifica 1, 00133 Roma, Italy}
\author{Massimo Bassan}
\affiliation{Dipartimento di Fisica,  Universit\`a di Tor Vergata, Via della Ricerca Scientifica 1, 00133  Roma, Italy}
\affiliation{Istituto Nazionale di Fisica Nucleare (INFN), Sezione di Roma Tor Vergata\\Via della Ricerca Scientifica 1, 00133 Roma, Italy}
\author{Giuseppe Pucacco}
\affiliation{Dipartimento di Fisica,  Universit\`a di Tor Vergata, Via della Ricerca Scientifica 1, 00133  Roma, Italy}
\affiliation{Istituto Nazionale di Fisica Nucleare (INFN), Sezione di Roma Tor Vergata\\Via della Ricerca Scientifica 1, 00133 Roma, Italy}
\author{Carmen Pardini}
\affiliation{Istituto di Scienza e Tecnologie dell'Informazione (ISTI) - Consiglio Nazionale delle Ricerche (CNR)\\ Via Moruzzi 1, 56124 Pisa, Italy}
\author{Luciano Anselmo}
\affiliation{Istituto di Scienza e Tecnologie dell'Informazione (ISTI) - Consiglio Nazionale delle Ricerche (CNR)\\ Via Moruzzi 1, 56124 Pisa, Italy}
\author{Carmelo Magnafico}
\affiliation{Istituto di Astrofisica e Planetologia Spaziali (IAPS) - Istituto Nazionale di Astrofisica (INAF)\\Via del Fosso del Cavaliere 100, 00133 Roma, Italy}
\affiliation{Istituto Nazionale di Fisica Nucleare (INFN), Sezione di Roma Tor Vergata\\Via della Ricerca Scientifica 1, 00133 Roma, Italy}
\date{\today}

\begin{abstract}
We present a new measurement of the Lense-Thirring effect on the orbits of the geodetic satellites LAGEOS, LAGEOS II and LARES. This secular precession is a general relativity effect produced by the gravitomagnetic field of the Earth generated by its rotation. The effect is a manifestation of spacetime curvature generated by mass-currents, a peculiarity of Einstein's theory of gravitation. This measurement stands out, compared to previous measurements in the same context, for its precision (\(\simeq7.4\times10^{-3}\)) and accuracy (\(\simeq16\times10^{-3}\)),
 i.e. for a reliable and robust evaluation of the systematic sources of error due to both gravitational and non-gravitational perturbations.
For this new measurement, we have largely exploited the results of GRACE mission to significantly improve the description of the gravitational field of the Earth, by also modeling its time dependence.
In this way, we strongly reduced the systematic errors due to the uncertainty in the knowledge of the Earth even zonal harmonics and, at the same time, avoided a possible bias of the final result and, consequently, of the precision of the measurement, linked to a non-reliable handling of the unmodeled and mismodeled periodic effects. 
\end{abstract}


\maketitle

\textit{Introduction}.---The precession of the orbital plane of an Earth-bound satellite caused by the so-called Lense-Thirring (LT) effect \cite{Thirring1918b,1918Lense-Thirring}  represents one of the most peculiar predictions of Einstein's theory of general relativity (GR). The orbital plane of the satellite behaves as a sort of gyroscope dragged by the gravitomagnetic (GM) field produced by the angular momentum of the central (primary) body.
Gravitoelectromagnetism represents, in the weak-field and slow-motion (WFSM) limit of GR, a phenomenon formally analogous to classical electromagnetism. This limit implies: \({GM_{\oplus}}/{rc^2}\ll 1\), \({J_{\oplus}}/{M_{\oplus}rc}\ll 1\) and \({v}/{c} \ll 1\), where \(G\), \(M_{\oplus}\) and \(J_{\oplus}\) represent, respectively, the gravitational constant, the mass of the primary and its intrinsic angular momentum, \(c\) and \(v\) are the speed of light and that of the satellite, while \(r\) represents the satellite distance from the primary.
Indeed, we have a gravitoelectric field  \(\textbf{E}_{\text{GE}}\) produced by masses, anologous to the electric field produced by charges, and a gravitomagnetic field  \(\textbf{B}_{\text{GM}}\)  produced by mass-currents, analogous to the magnetic field produced by electric currents.
Consequently, GR predicts a formal counterpart to the Amp\'ere's law of magnetism, 
that has no analogue in Newton's theory of gravitation.
Gravitomagnetism describes the spacetime curvature produced by mass-currents, i.e. the effects that derive from the non-diagonal components of Riemmann curvature tensor.
Gravitomagnetism~\cite{1988nznf.conf..573T} has several important consequences linked to the origin of inertia in GR~\cite{1995grin.book.....C}: in cosmology, in relationship with Mach's Principle~\cite{Mach:1883,1916PhyZ...17..101E,1916AnP...354..769E}, and in the astrophysics of compact objects~\cite{Thorne1983}, where it takes part in explaining astrophysical phenomena of high energy, as powerful source in creating the accretion disk and the jets observed in quasars and active galactic nuclei, due to the presence of a rotating supermassive black hole in their centers  \cite{1978PhRvD..18.3598D,1978PhRvD..17.1518D,1982MNRAS.198..345M,1986Sci...234..224T,1988nznf.conf..573T}.
However, to this date we lack an observational direct evidence of the gravitomagnetic effects in a strong field regime. 
On the other hand, in the WFSM limit of GR, several different works have  successfully provided a measurement of the GM field produced by the Earth~\cite{1996NCimA.109..575C,2004Natur.431..958C,2006NewA...11..527C,2011PhRvL.106v1101Em,2016EPJC...76..120Cb,2017mas..conf..131Lb}. All these measurements had however a number of systematic errors that limited their precision well above the target value of \(1\%\).
In the case of a gyroscope around the Earth, the LT precession with respect to the asymptotic reference frame of distant stars can be written as:
\begin{equation}
\dot{\mathbf{\Omega}}_\text{LT} =
-\frac{1}{2c}\mathbf{B}_\text{GM} = \frac{G}{c^2r^3}\left[  3 (\mathbf{J}_{\oplus}\cdot \hat{\mathbf{r}})\hat{\mathbf{r}}-\mathbf{J}_{\oplus}\right].
  \label{eq:pre_LT}
\end{equation}
This result is known as \emph{``dragging of the gyroscopes''}, and also as Schiff effect~\cite{1960PhRvL...4..215S,1960PNAS...46..871S}. 
This direct effect on an orbiting gyroscope has been measured by the NASA and 
Stanford University space mission Gravity Probe B (GP-B).
GP-B has provided a measurement of frame-dragging with an accuracy of about \(19\%\)~\cite{2011PhRvL.106v1101Em,2015CQGra..32v4001Em}, after an intense and extended analysis effort, needed to remove unexpected disturbances due to electric patch effects, which prevented the team to reach the initial goal of a \(1\%\) measurement of the relativistic effect.
Conversely, in the case of the measurements performed so far with the LAGEOS satellites~\cite{2004Natur.431..958C,2006NewA...11..527C,2007AdSpR..39..324L}, and more recently also with the inclusion of the LARES satellite~\cite{2016EPJC...76..120Cb,2017mas..conf..131Lb}, the main limitation in the  {accuracy} of the measurement is provided by the uncertainties in the knowledge (and modeling) of the even zonal harmonic coefficients of the Earth's gravitational potential.  The orbital plane of the satellite behaves, in principle, as an external inertial frame not bound to the Earth and dragged by its rotation; such dragging is observable if the classical disturbances due to the main gravitational and non-gravitational perturbations are properly modeled during the orbit determination or can be removed \textit{a posteriori} or considered negligible on the time span of the orbits analysis.
The dragging effect manifest itself in a secular shift of the right-ascension of the ascending node (RAAN) \(\Omega\) and of the argument pericenter \(\omega\) of the considered satellites~\cite{1918Lense-Thirring}:
\begin{equation}
\label{eq:Omega_sec_LT2}
\dot{\Omega}_{\text{LT}} = \mu\frac{2G{J}_{\oplus}}{c^2a^3(1-e^2)^{3/2}}=\mu K_\text{LT},
\end{equation}
\begin{equation}
\label{eq:omega_sec_LT2}
\dot{\omega}_{\text{LT}} = - \mu\frac{6G{J}_{\oplus}}{c^2a^3(1-e^2)^{3/2}}\cos i=-3\mu K_\text{LT}\cos i,
\end{equation}
where \(i\) represents the inclination of the orbit of the satellite with respect to the equatorial plane of the Earth, \(a\) its semi-major axis and \(e\) the eccentricity. Finally, the dimensionless coefficient \(\mu\) represents the LT effect parameter, with \(\mu=1\) if GR is the correct theory of gravitation, and \(\mu=0\) in Newtonian physics.
  Tables \ref{tab:elementi} and \ref{tab:precessioni}  show, respectively, the mean orbital elements of the satellites and the expected relativistic LT precession on their orbits.

\begin{table}[h!]
  \caption[]{Mean orbital elements of LAGEOS, LAGEOS II and LARES.}
  \centering
  \begin{tabular}{@{}cccrrr@{}}
  \hline
     \hline
\rm{Element}   & Unit   & Symbol 		&\rm{LAGEOS}     & \rm{LAGEOS II} & \rm{LARES} \\
\hline
 semi-major axis \, & [{\rm km}] & {\it a}                  			& 12 270.00          & 12 162.07 		&   7 820.31     \\
 eccentricity  & & {\it e}                                 				&  0.0044          & 0.0138       	&   0.0012 \\
 inclination \, & [{\rm deg}] & \it {i}                           			&  109.84              & 52.66       		& 69.49  \\
\hline
\hline
\end{tabular}
\label{tab:elementi}
\end{table}
%
These mean Keplerian elements have been determined by a dedicated analysis of the orbits of the satellites~\cite{Lucchesietal2015b}. 
The {accuracy} of the previously cited measurements of the LT effect with laser-ranged satellites, was estimated in the range \(5\%-10\%\), but gave rise to a vibrant debate in the literature~\cite{2003CeMDA..86..277I,2005NewA...10..616I,2005IJMPD..14.1989L,2006NewA...11..527C,2017EPJC...77...73I,2018EPJC...78..880Cb}.
\begin{table}[h!]
  \caption[]{Rate, in millisecond of arc per year (mas/yr), for the secular Lense-Thirring precession on the right-ascension of the ascending node and on the argument of pericenter of LAGEOS, LAGEOS II and LARES satellites.}
  \centering
  \begin{tabular}{@{}cccc@{}}
  \hline
  \hline
\rm{Rate in the element}     	& \rm{LAGEOS}     	& \rm{LAGEOS II} 	& \rm{LARES} \\
\hline
\(\dot{\Omega}_{{\textrm{LT}}}\) & +30.67 & +31.50 & +118.48\\
\(\dot{\omega}_{{\textrm{LT}}}\) & +31.23 & \(-57.31\) & \(-334.68\)\\
\hline
\hline
\end{tabular}
\label{tab:precessioni}
\end{table}
In this Letter we present and describe a precise and {accurate} measurement of the LT precession by the LARASE (LAser RAnged Satellites Experiment) Team~\cite{Lucchesietal2015b}.
One of our key goals consisted in improving the modeling of the orbits of the three satellites, with the ultimate target of refining the results that can be achieved in testing gravitation~\cite{7180629b,2017mas..conf..131Lb,2019Univ....5..141L}.
These improvements concern the modeling of both gravitational and non-gravitational perturbations~\cite{LucchesiEGU2017p2b,2018CeMDA.130...66P,LucchesiEGU2019p2,Lucchesietal2015b,2016AdSpR..57.1928V,2017AcAau.140..469P,2018PhRvD..98d4034V,PardiniEGU2019p2b,2019Univ....5..141L}.

\textit{On the accuracy of the even zonal harmonics knowledge}.---The gravitational potential \(V\) of the Earth is usually expanded in terms of spherical harmonics, to account for the non-uniform distribution of the mass of our planet. If we restrict our attention to the even zonal harmonics coefficients, \(\bar{C}_{\ell,0}\) (i.e. those with degree \(\ell=\text{even}\) and order \(m=0\)), the geopotential may be written as~\cite{1959AJ.....64..367K,1966tsga.book.....K}:
%
\begin{equation}
V=-\frac{GM_{\oplus}}{r}\sum_{\ell=2}^{\infty}\bigg(\frac{R_{\oplus}}{r}\bigg)^{\ell}P_{\ell 0}(\sin \varphi)\bar{C}_{\ell,0},\label{eq:multipole}
\end{equation}
%
where \(\varphi\) is the latitude, \(R_{\oplus}\) and \(r\) are, respectively, the mean equatorial radius of the Earth and the geocentric distance, while \(P_{\ell 0}(\sin \varphi)\) are the Legendre polynomials.
%
The deviation from the spherical symmetry for the mass distribution of the Earth, described by these harmonics, is responsible for a classical precession of the satellite orbit both in the RAAN, \(\dot{\Omega}_{class}\), and in the argument of pericenter, \(\dot{\omega}_{class}\), just as it happens due to the LT precession, but with  much larger amplitudes.
For instance,  the  secular effect in the rate of the  RAAN due to the first two even zonal harmonics is:
%
%
\begin{widetext}
\begin{equation}
\label{eq:pre_class}
\dot{\Omega}_{\text{class}} = \frac{3}{2}n\left(\frac{R_{\oplus}}{a}\right)^2\frac{\cos i}{\left(1-e^2\right)^2} \left\lbrace \sqrt{5}\bar{C}_{2,0} +\frac{15}{8}\left(\frac{R_{\oplus}}{a}\right)^3(7\sin^2i -4) \frac{(1+\frac{3}{2}e^2)}{(1-e^2)^2}\bar{C}_{4,0}\right\rbrace = K_{2,0}\bar{C}_{2,0}+K_{4,0}\bar{C}_{4,0},
\end{equation}
\end{widetext}

where \(\bar{C}_{2,0}\) and \(\bar{C}_{4,0}\) represent the normalized Stokes coefficients of the quadrupole and octupole moments of the Earth and \(n\) the satellite mean motion~\footnote{The order-of-magnitude of this \textit{classical} precession is about \(+126\) deg./yr for LAGEOS, \(-231\) deg./yr for LAGEOS II, and about \(-624\) deg./yr in the case of LARES.}.
Thus, an imperfect knowledge of the Earth's multipolar moments can produce a large systematic error in the measurement of the LT precession. This aspect is further aggravated by the time dependence of these coefficients, that are characterized by several periodic effects  with, mainly, annual and inter-annual periodicities~\cite{1997JGR...10222377C,2002Sci...297..831C,JGRB:JGRB50058,2018GeoJI.212.1218C}.
To avoid a faulty measurement of the LT precession, we need to better account for this time dependence.  First steps in this direction were taken 
in \cite{2019Univ....5..141L} in the case of the quadrupole coefficient that is responsible for the largest effect; here, we extend those actions to multipole coefficients of higher order, at least up to degree \(\ell=20\)~\footnote{In \cite{2019Univ....5..141L}, we fitted linearly the quadrupole coefficient obtained from GRACE (Gravity Recovery And Climate Experiment) monthly solutions, and we used this fitted value in the data reduction of the satellites orbit. Furthermore, in that paper, and for the GGM05S model, with regard to the LT effect measurement, we have also analyzed  and compared the error related to the knowledge of the octupole coefficient with respect to the hexapole one.}.


\textit{New aspects}.--- This improvement was made possible by the new solutions of the Earth's gravitational field provided, especially, by the GRACE  space mission.
The twin GRACE satellites, launched on March 2002 by NASA and DLR, have provided a much better determination of the Earth's gravitational field, both in its static and dynamical components \cite{2002AdSpR..30..129R,2003AdSpR..31.1883Rb,2005JGeo...39....1Rb,2001AGUFM.G41C..02T},  with respect to previous results based on multi-satellite data.
Moreover, the correlations among various coefficients, in particular  the even zonal ones, have been reduced with respect to previous models, such as EGM96 ~\cite{1998Lemoineb}.

In our analyses, and this represents a first element of novelty, we considered the monthly solutions for these coefficients provided by three independent analysis centers: CSR,  GFZ, and JPL~\footnote{See the International Centre for Global Earth Models (ICGEM): Gravity Field Solutions for dedicated Time Periods: Release 05, http://icgem.gfz-potsdam.de/series (2018).}.
We then fitted these coefficients with a linear trend on the time span of our analysis, about 6.5 years, and we modeled each even zonal harmonics between \(\ell=2\) and  \(\ell=20\) following such behaviour~\footnote{This was also extended up to degree \(\ell=30\), but with no appreciable difference in the results. In the case of the quadrupole coefficient we also considered a more complete non-linear fit to the time behaviour outlined by GRACE, with the inclusion of two periodic terms, one with a yearly frequency and one at twice this frequency. However, the final results have not shown a noticeable difference with respect to those obtained with a simpler linear fit.}.
In all previous measurements of the LT effect, only the quadrupole and octupole coefficients where modeled, by means of a linear trend~\cite{2010IERS-Conv-2010}~\footnote{Usually, these trends were those suggested by IERS Conventions~\cite{2010IERS-Conv-2010}, but were not always compatible with the results from GRACE monthly solutions.}.
The linear trend \textit{``follows''}, in average, the more complex time dependence of these coefficients, that are also characterized by the mentioned annual and semi-annual periodicities, plus other subtler periodic effects~\cite{2013GeoRL..40.2625C,JGRB:JGRB50058}.
This approach strongly reduces the discrepancies between the effective time behaviour of these coefficients and their static values provided in several models~\footnote{It is important to stress that the static models provide, in general, good measurements for the coefficients of the Earth's gravity field on the time span of GRACE data over which they have been effectively computed, i.e. they represent averages values on this time interval. However, on different subintervals of the entire period of GRACE data, as in the case of our analysis, that starts after the launch of the LARES satellites, these average values may be quite different with respect to their effective time behaviour provided by GRACE monthly solutions.}.

As usual practice in this kind of analysis, we estimated, together with the relativistic precession, the corrections to the quadrupole \(\bar{C}_{2,0}\) and octupole \(\bar{C}_{4,0}\) coefficients, (by solving Eqs. \ref{eq:sistema} below); these are the coefficients that most contribute to the systematic errors, if not properly accounted for in the precise orbit determination (POD) of the satellites. A second element of novelty is represented by the fact that we  also explicitly computed their correlation with the relativistic parameter.

A third element of novelty is represented by the adoption of different models for the background gravitational field of the Earth, while only one model was used in the past. This allows us, once the first 10 even zonal harmonics for each field were set in agreement with GRACE monthly solutions, to highlight the variability of the systematic errors among the considered fields. Consequently, this variability is due to even zonal harmonics of higher order, as well as to the non-zonal ones (tesseral and sectorial, respectively with \(m \neq \ell\) and with \(m=\ell\)).

\textit{The new analyses}.---We analysed the tracking data of LAGEOS, LAGEOS II and LARES on a time span of 2359 days (about 6.5 years), starting from April 6, 2012 (MJD 56023).
The GEODYN II \cite{1998pavlis,1990AdSpR..10..197P} software was used for the data reduction of the Satellite Laser Ranging (SLR) observations of the three satellites in the normal point (NP) format~\cite{Sinclair1997}. These PODs have been performed by adjusting the state-vector (i.e. position and velocity) of each satellite on an arc length of 7 days, for an overall number of 337 arcs not causally connected. The relativistic acceleration on the orbits of the satellites produced by Earth's rotation was not included in the dynamical model of GEODYN~\cite{1990CeMDA..48..167H}. Also the subtle thermal thrust effects due to Earth's infrared radiation (the Earth-Yarkovsky effect) and to Sun visible radiation, modulated by the eclipses (the solar Yarkovsky-Schach effect)~\cite{2002P&SS...50.1067L}), were not included in the dynamical model, due to their still inadequate modeling~\footnote{
This exclusion is motivated, due to the complexity of the effects and their dependence on the spin vector evolution of the satellite, by the fact that the routines included in GEODYN for their modelling are not updated. Moreover they were valid only in the fast rotation regime of the satellites, which is not longer applicable for the two LAGEOS satellites \cite{2013AdSpR..52.1332K,2018PhRvD..98d4034V}. 
For the same reason we had to exclude the pericenter from our analysis. Our ongoing effort to improve the model for the thermal thrust perturbations~\cite{2002P&SS...50.1067L,2016AdSpR..57.1928V,2018PhRvD..98d4034V,2019Univ....5..141L} will hopefully allow us to include, in forthcoming measurements, also this element in the analysis, especially for  LAGEOS II.}.

In our new analyses we used  the GGM05S solution~\cite{Tapley2013,JGRB:JGRB50058} as reference model for the Earth's gravitational field~\footnote{This is the solution currently used by the International Laser Ranging Service (ILRS) to realize the International Terrestrial Reference Frame (ITRF2014) \cite{2016JGRB..121.6109A}, i.e. the practical realization of the International Terrestrial Reference System (ITRS)~\cite{2010IERS-Conv-2010}.}.
Other models considered in this work are: EIGEN-GRACE02S (2004) \cite{2005JGeo...39....1Rb}, \verb|ITU_GRACE16| (2016) \cite{ITU-GRACE16b} and Tonji-Grace02s (2017) \cite{Tongji-Grace02s,2018JGRB..123.6111C}.
All these models were obtained from GRACE data.

Concerning tides, we included solid tides with Colombo's model \cite{1984aot..book.....C}, while ocean tides were described by GOT99.2 model \cite{1999Ray}. Furthermore, with regard to the on-ground tracking stations, the ILRS recommendations for the time span of our analyses have been followed~\cite{ilrs2}. 
We remark that the time span of our analyses for the orbits of the three satellites, about 6.5 years, is close to twice the period of the node of the older LAGEOS (1052 days), 4 times the period of the node of LAGEOS II (570 days) and 11 times that of LARES (211 days).
This allows us to reduce the impact on the relativistic measurement 
of all the unmodeled (or poorly modeled) effects related to the non-gravitational and gravitational perturbations that are characterized by these periodicities. For instance, this is the case for some thermal thrust perturbations~\cite{2002P&SS...50.1067L} or of the long-period ocean tide \(K_1\)~\cite{2018CeMDA.130...66P}. 
In the following, the most significant results we have obtained from the analysis of the residuals in the RAAN of the three satellites are described~\footnote{We have estimated the LT effect from the orbit of LAGEOS, LAGEOS II and LARES applying several different strategies. Some of these correspond to new methods with respect to those reported in the published literature on the LT effect measurements. We cannot present all these methods in this Letter, but we can disclose that the final results obtained for \(\mu\) are all consistent, with a precision at the level of a fraction of percent for the cumulative residuals.}.

\begin{figure}[h!]
	\centering
	\includegraphics[width=0.3\textwidth]{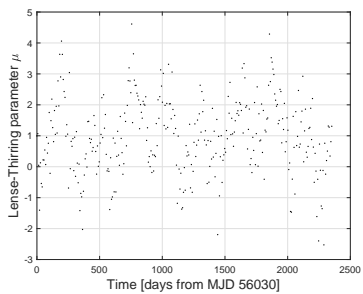}
	\caption{The Lense-Thirring parameter \(\mu\) estimated, arc by arc, on the time span of our analysis. For the mean value we obtained \(\mu=0.99\pm0.13\), with a standard deviation \(\sigma\) of about 1.20. This large value of \(\sigma\) is due to the unmodeled periodic effects.}
	\label{fig:mu}
\end{figure}
\begin{figure}[h!]
	\centering
	\includegraphics[width=0.3\textwidth]{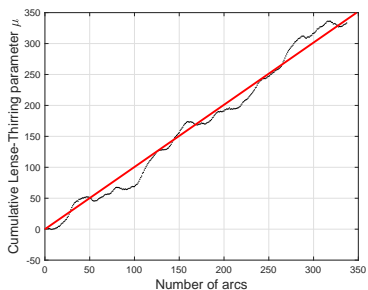}
	\caption{The Lense-Thirring parameter \(\mu\) (black dots) obtained as the slope of the cumulative sum of the values shown in Fig. \ref{fig:mu}. The horizontal axis shows the number of arcs of the satellites POD. 
	The slope of the linear fit (red line) provides a value of \(\mu = 1.0053\pm 0.0074\), see Table \ref{tab:gravity-fields}. The impact of the unmodeled periodic effects is reduced in the case of the cumulative residuals.}
	\label{fig:mu_int}
\end{figure}

\textit{Results}.---For each of the considered gravitational models, where however the first 10 even zonal harmonics were inserted with the time behaviour described above, 
we have solved (arc by arc) a system of three equations in the three unknowns {\(\delta \bar{C}_{2,0}\), \(\delta \bar{C}_{4,0}\) and \(\mu\): }
\begin{eqnarray}
\left\{
\begin{array}{c c c c c}
K_{2,0}^{L1}\delta \bar{C}_{2,0} + K_{4,0}^{L1}\delta \bar{C}_{4,0} +  K_{LT}^{L1}\mu = \delta\dot{\Omega}^{L1}_{res} \\
K_{2,0}^{L2}\delta \bar{C}_{2,0} + K_{4,0}^{L2}\delta \bar{C}_{4,0} +  K_{LT}^{L2}\mu = \delta\dot{\Omega}^{L2}_{res} \\
\label{eq3}
K_{2,0}^{LR}\delta \bar{C}_{2,0} + K_{4,0}^{LR}\delta \bar{C}_{4,0} +  K_{LT}^{LR}\mu = \delta\dot{\Omega}^{LR}_{res},
\end{array}
\right.
\label{eq:sistema}
\end{eqnarray}
following the original proposal of~\cite{1996NCimA.109.1709C}.
In these equations~\footnote{In this system of Eqs. we are neglecting the contributions from the mismodeling of the higher harmonics. Anyway, their contribution is explicitly considered in the evaluation of the systematic errors, i.e. in the overall error budget of the measurement.}, the constant terms \(\delta\dot{\Omega}^{L1}_{res}\), \(\delta\dot{\Omega}^{L2}_{res}\) and \(\delta\dot{\Omega}^{LR}_{res}\) represent the residuals in the rate of the RAAN of each satellite determined, after the POD of their orbit, with the method described in \cite{2006P&SS...54..581L}.
In the left hand side of each equation,  the coefficients \(K^{Lj}_{2,0}\) and \(K^{Lj}_{4,0}\)
 (with \(Lj = (L1)\text{LAGEOS}, (L2)\text{LAGEOS II}, (LR)\text{LARES}\)) are function of the orbital parameters \(a\), \(e\) and \(i\) of each satellite, as explicitly shown in Eq. (\ref{eq:pre_class}) ~\footnote{The orbital parameters are known with a very small relative uncertainty, such that the \(K\) coefficients can be considered, for our purposes, as error-free.}.
The three coefficients \(K_{LT}^{Lj}\) represent the relativistic Lense-Thirring precession on the three satellites, see Eq. (\ref{eq:Omega_sec_LT2}) and Table \ref{tab:precessioni}. 
Finally, concerning the three unknowns of the system,
 the quantities \(\delta \bar{C}_{\ell,0}\) (with \(\ell=2\) and \(\ell=4\)) represent the mismodeling of the even zonal harmonics of the background model of the Earth's gravitational field, while \(\mu\) represents the dimensionless LT parameter introduced in Eq. (\ref{eq:Omega_sec_LT2})~\footnote{In this Letter we directly provide  the result for \(\mu\) and not, as done in previous measurements of the LT effect, for the combined rate of the RAAN of the three satellites. In our analyses, this combination corresponds to a precession of 50.17 mas/yr.}.
Therefore, with this solution we can remove, from the estimate of \(\mu\), the uncertainties related to the knowledge of the first two even zonal harmonics.
In Fig. \ref{fig:mu} we show the results for the LT parameter \(\mu\),  evaluated, for each arc, as a solution of the system of Eqs. (\ref{eq:sistema}). The set shown was computed using the GGM05S Earth model.
In  Fig. \ref{fig:mu_int},  \(\mu\) is plotted as a cumulative sum of the values shown in Fig. \ref{fig:mu}.

To extract the LT parameter, we compute the mean value in the case of Fig. \ref{fig:mu}, and the slope of a best fitting straight line in the case of Fig. \ref{fig:mu_int}. 
It can be seen that periodic effects perturb, especially in the case of Fig. \ref{fig:mu}, the measurement of the relativistic secular effect embedded in the time series of the residuals.
We underline, however, that the unmodeled periodic effects that characterize the residuals  behave  ---  in relation of their peculiar nature that alternatively increases or decreases the unmodeled precession on various arcs --- 
as a Gaussian distribution superimposed to the relativistic precession.
Indeed, the distribution of the combined residuals of Fig. \ref{fig:mu}  is very close to a Gaussian, with a Skewness of \(-8.4\times10^{-3}\) and a Kurtosis of  \(+3.097\).
Since the cumulative sum acts as a low-pass filter, the residuals shown in Fig. \ref{fig:mu_int} are less influenced by the high frequencies present in Fig. \ref{fig:mu}, and allow for a  \textit{cleaner}, and consequently more precise, measurement of the LT effect.
In Table \ref{tab:corr} we show, still in the case of the GGM05S model, the correlation matrix for the three solutions of the system of Eqs. (\ref{eq:sistema}).

\begin{table}[h!]
  \caption[]{Correlations among the three estimated quantities evaluated as solutions of Eqs.  (\ref{eq:sistema}) in the case of the GGM05S model: the correction \(\delta\bar{C}_{2,0}\) to the quadrupole coefficient, the correction \(\delta\bar{C}_{4,0}\) to the octupole coefficient and the Lense-Thirring parameter \(\mu\).}
  \centering 
  \begin{tabular}{@{}cccc@{}}
  \hline
  \hline
    	& \rm{\(\delta\bar{C}_{2,0}\)}     	& \rm{\(\delta\bar{C}_{4,0}\)}  	& \rm{\(\mu\)}  \\
\hline
\(\delta\bar{C}_{2,0}\) & +1.000 & +0.082 & +0.071\\
\(\delta\bar{C}_{4,0}\) & +0.082 & +1.000 & \(-0.179\)\\
\(\mu\) & +0.071 & \(-0.179\) & +1.000\\
\hline
\hline
\end{tabular}
\label{tab:corr}
\end{table}
The correlation is particularly small between \(\mu\) and the correction \(\delta\bar{C}_{2,0}\) to the quadrupole coefficient, as well as between the two corrections, \(\delta\bar{C}_{2,0}\) and \(\delta\bar{C}_{4,0}\), estimated for the two lowest even zonal harmonics.
Conversely, the correlation is moderately larger between the Lense-Thirring parameter and the correction \(\delta\bar{C}_{4,0}\) to the octupole coefficient. 
Similar results  are found when using all other gravitational fields considered.
These analyses show that  the correlations among the estimated quantities  can be reduced, with respect to a previous result~\cite{2019Univ....5..141L}, where  the LT effect was estimated together with the corrections to the quadrupole and hexapole, \(\delta\bar{C}_{6,0}\), coefficients.
In Table \ref{tab:gravity-fields} we show the resulting value for the LT parameter \(\mu\), estimated from the slope of the linear fit to the cumulative residuals, for each of the models considered in our analyses. 

\begin{table}[h!]
  \caption[]{Comparison of the results for the measurement of the relativistic LT precession among different models for the background  gravitational field of the Earth, computed with the method of cumulative residuals.
Column 1 provides the model considered. Column 2 provides the measured slope with its \(2\sigma\) error (errors appear to be equal, due to rounding), i.e. the LT parameter \(\mu\) for the unmodeled secular effect. These results for the measured parameter \(\mu\) are compatible (within their errors) with the prediction of GR for the relativistic precession, i.e. with  \(\mu_{GR}=1\).}
  \centering
  \begin{tabular}{@{}cc@{}}
  \hline
  \hline
\multicolumn{1}{c}{\rm{Model}}  &  \multicolumn{1}{c}{\(\mu \pm \delta\mu\)} \\
\hline
GGM05S &  \(1.0053\pm 0.0074\)  \\
EIGEN-GRACE02S & \(1.0002\pm 0.0074\)  \\
ITU\_GRACE16 &    \(0.9996\pm 0.0074\)  \\
Tonji-Grace02s   &  \(1.0008\pm 0.0074\)  \\
\hline
\hline
\end{tabular}
\label{tab:gravity-fields}
\end{table}
We notice that, in all these analyses, the agreement with the GR prediction is at a fraction of  percent. Also the precision \(\delta\mu\) (at a \(2\sigma\) level) is at a fraction of percent of the relativistic precession. As already highlighted, the small variability of the results is mainly related to possible systematic differences among the different solutions for the gravitational field, due to  even zonal harmonic coefficients with degree \(\ell > 20\).

Assuming, conservatively, for this measurement of the LT effect, the mean of the measured values of \(\mu\)  reported in Table \ref{tab:gravity-fields}, we obtain \(\mu_\text{meas}=1.0015\). Considering also the errors from the fit and an independent evaluation of the main systematic sources of error, we obtain the following result:
\begin{equation}
{\mu}_\text{meas} - 1 = (1.5 \pm 7.4)\times10^{-3} \pm 16\times10^{-3}. \label{eq:_LT_meas_sys}
\end{equation}
The error budget of about \(1.6\%\) derives from a root-sum-square of the the main systematic effects related to the gravitational and non-gravitational perturbations that act on the orbits of the satellites. It accounts for the errors related to the static field (\(\simeq 1.0\%\)), to ocean tides (\(\lesssim 0.6\%\)), to other periodic effects (\(\simeq 1.0\%\)) and to the error related to the knowledge of the de Sitter precession (\(\simeq 0.3\%\)).
The details of the error budget will be presented in a more extended paper, with a complete discussion of the results obtained with the various approaches we have considered, including the results for the measurement of the LT effect in the case of the non-cumulative residuals. 

This result represents a precise and {accurate} measurement of the Lense-Thirring effect in the field of the Earth with laser-ranged satellites, where many of the criticalities raised in the past have been overcome. Indeed, several new improvements have been introduced. Among these, we believe that the most significant is the modeling in the POD of the even zonal harmonics through linear trends that fit the monthly solutions of GRACE coefficients on the time interval of our analyses. Moreover, we also estimated, together with the relativistic parameter \(\mu\), the corrections to the quadrupole and octupole coefficients and the correlation values among these quantities.
These improvements have allowed us to perform the measurement by means of a simple linear fit to the orbit residuals. 
Indeed, whenever some of the  coefficients of the gravitational field are not correctly modeled in the POD, the simple linear fit preserves the analysis from the risk of forcing the result to the desired value, using an excessive number of fitting parameters: since these could absorb some of the periodic effects beyond our ability of understanding their origin.

The authors acknowledge the ILRS for providing high quality laser ranging data of the two LAGEOS satellites and of LARES. This work has been performed by the LARASE experiment and funded by the Commissione Scientifica Nazionale (CSN2) for Astroparticle Physics experiment of the Istituto Nazionale di Fisica Nucleare (INFN), to which we are very grateful.

\bibliography{LARASE,LARASE_universe}

\begin{thebibliography}{76}%
\makeatletter
\providecommand \@ifxundefined [1]{%
 \@ifx{#1\undefined}
}%
\providecommand \@ifnum [1]{%
 \ifnum #1\expandafter \@firstoftwo
 \else \expandafter \@secondoftwo
 \fi
}%
\providecommand \@ifx [1]{%
 \ifx #1\expandafter \@firstoftwo
 \else \expandafter \@secondoftwo
 \fi
}%
\providecommand \natexlab [1]{#1}%
\providecommand \enquote  [1]{``#1''}%
\providecommand \bibnamefont  [1]{#1}%
\providecommand \bibfnamefont [1]{#1}%
\providecommand \citenamefont [1]{#1}%
\providecommand \href@noop [0]{\@secondoftwo}%
\providecommand \href [0]{\begingroup \@sanitize@url \@href}%
\providecommand \@href[1]{\@@startlink{#1}\@@href}%
\providecommand \@@href[1]{\endgroup#1\@@endlink}%
\providecommand \@sanitize@url [0]{\catcode `\\12\catcode `\$12\catcode
  `\&12\catcode `\#12\catcode `\^12\catcode `\_12\catcode `\%12\relax}%
\providecommand \@@startlink[1]{}%
\providecommand \@@endlink[0]{}%
\providecommand \url  [0]{\begingroup\@sanitize@url \@url }%
\providecommand \@url [1]{\endgroup\@href {#1}{\urlprefix }}%
\providecommand \urlprefix  [0]{URL }%
\providecommand \Eprint [0]{\href }%
\providecommand \doibase [0]{http://dx.doi.org/}%
\providecommand \selectlanguage [0]{\@gobble}%
\providecommand \bibinfo  [0]{\@secondoftwo}%
\providecommand \bibfield  [0]{\@secondoftwo}%
\providecommand \translation [1]{[#1]}%
\providecommand \BibitemOpen [0]{}%
\providecommand \bibitemStop [0]{}%
\providecommand \bibitemNoStop [0]{.\EOS\space}%
\providecommand \EOS [0]{\spacefactor3000\relax}%
\providecommand \BibitemShut  [1]{\csname bibitem#1\endcsname}%
\let\auto@bib@innerbib\@empty
\bibitem [{\citenamefont {{Thirring}}(1918)}]{Thirring1918b}%
  \BibitemOpen
  \bibfield  {author} {\bibinfo {author} {\bibfnamefont {H.}~\bibnamefont
  {{Thirring}}},\ }\href@noop {} {\bibfield  {journal} {\bibinfo  {journal}
  {Physikalische Zeitschrift}\ }\textbf {\bibinfo {volume} {19}},\ \bibinfo
  {pages} {33} (\bibinfo {year} {1918})}\BibitemShut {NoStop}%
\bibitem [{\citenamefont {{Lense}}\ and\ \citenamefont
  {{Thirring}}(1918)}]{1918Lense-Thirring}%
  \BibitemOpen
  \bibfield  {author} {\bibinfo {author} {\bibfnamefont {J.}~\bibnamefont
  {{Lense}}}\ and\ \bibinfo {author} {\bibfnamefont {H.}~\bibnamefont
  {{Thirring}}},\ }\href@noop {} {\bibfield  {journal} {\bibinfo  {journal}
  {Phys. Z.}\ }\textbf {\bibinfo {volume} {19}},\ \bibinfo {pages} {156}
  (\bibinfo {year} {1918})}\BibitemShut {NoStop}%
\bibitem [{\citenamefont {{Thorne}}(1988)}]{1988nznf.conf..573T}%
  \BibitemOpen
  \bibfield  {author} {\bibinfo {author} {\bibfnamefont {K.~S.}\ \bibnamefont
  {{Thorne}}},\ }in\ \href@noop {} {\emph {\bibinfo {booktitle} {{Near Zero:
  New Frontiers of Physics}}}},\ \bibinfo {editor} {edited by\ \bibinfo
  {editor} {\bibfnamefont {J.~D.}\ \bibnamefont {{Fairbank}}}, \bibinfo
  {editor} {\bibfnamefont {J.~B.~S.}\ \bibnamefont {{Deaver}}}, \bibinfo
  {editor} {\bibfnamefont {C.~W.~F.}\ \bibnamefont {{Everitt}}}, \ and\
  \bibinfo {editor} {\bibfnamefont {P.~F.}\ \bibnamefont {{Michelson}}}}\
  (\bibinfo {year} {1988})\ pp.\ \bibinfo {pages} {573--586}\BibitemShut
  {NoStop}%
\bibitem [{\citenamefont {{Ciufolini}}\ and\ \citenamefont
  {{Wheeler}}(1995)}]{1995grin.book.....C}%
  \BibitemOpen
  \bibfield  {author} {\bibinfo {author} {\bibfnamefont {I.}~\bibnamefont
  {{Ciufolini}}}\ and\ \bibinfo {author} {\bibfnamefont {J.~A.}\ \bibnamefont
  {{Wheeler}}},\ }\href@noop {} {\emph {\bibinfo {title} {{Gravitation and
  inertia}}}}\ (\bibinfo  {publisher} {Princeton University Press},\ \bibinfo
  {address} {Princeton},\ \bibinfo {year} {1995})\BibitemShut {NoStop}%
\bibitem [{\citenamefont {Mach}()}]{Mach:1883}%
  \BibitemOpen
  \bibfield  {author} {\bibinfo {author} {\bibfnamefont {E.}~\bibnamefont
  {Mach}},\ }\href@noop {} {\emph {\bibinfo {title} {{Die Mechanik in ihrer
  Entwickelung historisch-kritisch dargestellt}}}}\BibitemShut {NoStop}%
\bibitem [{\citenamefont
  {{Einstein}}(1916{\natexlab{a}})}]{1916PhyZ...17..101E}%
  \BibitemOpen
  \bibfield  {author} {\bibinfo {author} {\bibfnamefont {A.}~\bibnamefont
  {{Einstein}}},\ }\href@noop {} {\bibfield  {journal} {\bibinfo  {journal}
  {Physikalische Zeitschrift}\ }\textbf {\bibinfo {volume} {17}},\ \bibinfo
  {pages} {101} (\bibinfo {year} {1916}{\natexlab{a}})}\BibitemShut {NoStop}%
\bibitem [{\citenamefont
  {{Einstein}}(1916{\natexlab{b}})}]{1916AnP...354..769E}%
  \BibitemOpen
  \bibfield  {author} {\bibinfo {author} {\bibfnamefont {A.}~\bibnamefont
  {{Einstein}}},\ }\href {\doibase 10.1002/andp.19163540702} {\bibfield
  {journal} {\bibinfo  {journal} {Annalen der Physik}\ }\textbf {\bibinfo
  {volume} {354}},\ \bibinfo {pages} {769} (\bibinfo {year}
  {1916}{\natexlab{b}})}\BibitemShut {NoStop}%
\bibitem [{\citenamefont {Thorne}(1983)}]{Thorne1983}%
  \BibitemOpen
  \bibfield  {author} {\bibinfo {author} {\bibfnamefont {K.~S.}\ \bibnamefont
  {Thorne}},\ }\enquote {\bibinfo {title} {{Quantum Optics, Experimental
  Gravity, and Measurement Theory}},}\ \ (\bibinfo  {publisher} {Springer US},\
  \bibinfo {address} {Boston, MA},\ \bibinfo {year} {1983})\ Chap.\ \bibinfo
  {chapter} {Experimental Gravity, Gravitational Waves, and Quantum
  Nondemolition; An Introduction}, pp.\ \bibinfo {pages} {325--346}\BibitemShut
  {NoStop}%
\bibitem [{\citenamefont {{Damour}}(1978)}]{1978PhRvD..18.3598D}%
  \BibitemOpen
  \bibfield  {author} {\bibinfo {author} {\bibfnamefont {T.}~\bibnamefont
  {{Damour}}},\ }\href {\doibase 10.1103/PhysRevD.18.3598} {\bibfield
  {journal} {\bibinfo  {journal} {\prd}\ }\textbf {\bibinfo {volume} {18}},\
  \bibinfo {pages} {3598} (\bibinfo {year} {1978})}\BibitemShut {NoStop}%
\bibitem [{\citenamefont {{Damour}}\ \emph {et~al.}(1978)\citenamefont
  {{Damour}}, \citenamefont {{Hanni}}, \citenamefont {{Ruffini}},\ and\
  \citenamefont {{Wilson}}}]{1978PhRvD..17.1518D}%
  \BibitemOpen
  \bibfield  {author} {\bibinfo {author} {\bibfnamefont {T.}~\bibnamefont
  {{Damour}}}, \bibinfo {author} {\bibfnamefont {R.~S.}\ \bibnamefont
  {{Hanni}}}, \bibinfo {author} {\bibfnamefont {R.}~\bibnamefont {{Ruffini}}},
  \ and\ \bibinfo {author} {\bibfnamefont {J.~R.}\ \bibnamefont {{Wilson}}},\
  }\href {\doibase 10.1103/PhysRevD.17.1518} {\bibfield  {journal} {\bibinfo
  {journal} {\prd}\ }\textbf {\bibinfo {volume} {17}},\ \bibinfo {pages} {1518}
  (\bibinfo {year} {1978})}\BibitemShut {NoStop}%
\bibitem [{\citenamefont {{MacDonald}}\ and\ \citenamefont
  {{Thorne}}(1982)}]{1982MNRAS.198..345M}%
  \BibitemOpen
  \bibfield  {author} {\bibinfo {author} {\bibfnamefont {D.}~\bibnamefont
  {{MacDonald}}}\ and\ \bibinfo {author} {\bibfnamefont {K.~S.}\ \bibnamefont
  {{Thorne}}},\ }\href {\doibase 10.1093/mnras/198.2.345} {\bibfield  {journal}
  {\bibinfo  {journal} {Mon. Not. Roy. Astron. Soc.}\ }\textbf {\bibinfo
  {volume} {198}},\ \bibinfo {pages} {345} (\bibinfo {year}
  {1982})}\BibitemShut {NoStop}%
\bibitem [{\citenamefont {{Thorne}}\ \emph {et~al.}(1986)\citenamefont
  {{Thorne}}, \citenamefont {{Price}},\ and\ \citenamefont
  {{MacDonald}}}]{1986Sci...234..224T}%
  \BibitemOpen
  \bibfield  {author} {\bibinfo {author} {\bibfnamefont {K.~S.}\ \bibnamefont
  {{Thorne}}}, \bibinfo {author} {\bibfnamefont {R.~H.}\ \bibnamefont
  {{Price}}}, \ and\ \bibinfo {author} {\bibfnamefont {D.~A.}\ \bibnamefont
  {{MacDonald}}},\ }\href@noop {} {\bibfield  {journal} {\bibinfo  {journal}
  {Science}\ }\textbf {\bibinfo {volume} {234}},\ \bibinfo {pages} {224}
  (\bibinfo {year} {1986})}\BibitemShut {NoStop}%
\bibitem [{\citenamefont {{Ciufolini}}\ \emph {et~al.}(1996)\citenamefont
  {{Ciufolini}}, \citenamefont {{Lucchesi}}, \citenamefont {{Vespe}},\ and\
  \citenamefont {{Mandiello}}}]{1996NCimA.109..575C}%
  \BibitemOpen
  \bibfield  {author} {\bibinfo {author} {\bibfnamefont {I.}~\bibnamefont
  {{Ciufolini}}}, \bibinfo {author} {\bibfnamefont {D.}~\bibnamefont
  {{Lucchesi}}}, \bibinfo {author} {\bibfnamefont {F.}~\bibnamefont {{Vespe}}},
  \ and\ \bibinfo {author} {\bibfnamefont {A.}~\bibnamefont {{Mandiello}}},\
  }\href {\doibase 10.1007/BF02731140} {\bibfield  {journal} {\bibinfo
  {journal} {Nuovo Cim. A}\ }\textbf {\bibinfo {volume} {109}},\ \bibinfo
  {pages} {575} (\bibinfo {year} {1996})}\BibitemShut {NoStop}%
\bibitem [{\citenamefont {{Ciufolini}}\ and\ \citenamefont
  {{Pavlis}}(2004)}]{2004Natur.431..958C}%
  \BibitemOpen
  \bibfield  {author} {\bibinfo {author} {\bibfnamefont {I.}~\bibnamefont
  {{Ciufolini}}}\ and\ \bibinfo {author} {\bibfnamefont {E.~C.}\ \bibnamefont
  {{Pavlis}}},\ }\href {\doibase 10.1038/nature03007} {\bibfield  {journal}
  {\bibinfo  {journal} {Nature}\ }\textbf {\bibinfo {volume} {431}},\ \bibinfo
  {pages} {958} (\bibinfo {year} {2004})}\BibitemShut {NoStop}%
\bibitem [{\citenamefont {{Ciufolini}}\ \emph {et~al.}(2006)\citenamefont
  {{Ciufolini}}, \citenamefont {{Pavlis}},\ and\ \citenamefont
  {{Peron}}}]{2006NewA...11..527C}%
  \BibitemOpen
  \bibfield  {author} {\bibinfo {author} {\bibfnamefont {I.}~\bibnamefont
  {{Ciufolini}}}, \bibinfo {author} {\bibfnamefont {E.~C.}\ \bibnamefont
  {{Pavlis}}}, \ and\ \bibinfo {author} {\bibfnamefont {R.}~\bibnamefont
  {{Peron}}},\ }\href {\doibase 10.1016/j.newast.2006.02.001} {\bibfield
  {journal} {\bibinfo  {journal} {New Astron.}\ }\textbf {\bibinfo {volume}
  {11}},\ \bibinfo {pages} {527} (\bibinfo {year} {2006})}\BibitemShut
  {NoStop}%
\bibitem [{\citenamefont {{Everitt}}\ \emph {et~al.}(2011)\citenamefont
  {{Everitt}}, \citenamefont {{Debra}}, \citenamefont {{Parkinson}},\ and\
  \citenamefont {et~al.}}]{2011PhRvL.106v1101Em}%
  \BibitemOpen
  \bibfield  {author} {\bibinfo {author} {\bibfnamefont {C.~W.~F.}\
  \bibnamefont {{Everitt}}}, \bibinfo {author} {\bibfnamefont {D.~B.}\
  \bibnamefont {{Debra}}}, \bibinfo {author} {\bibfnamefont {B.~W.}\
  \bibnamefont {{Parkinson}}}, \ and\ \bibinfo {author} {\bibnamefont
  {et~al.}},\ }\href {\doibase 10.1103/PhysRevLett.106.221101} {\bibfield
  {journal} {\bibinfo  {journal} {Physical Review Letters}\ }\textbf {\bibinfo
  {volume} {106}},\ \bibinfo {eid} {221101} (\bibinfo {year} {2011})},\ \Eprint
  {http://arxiv.org/abs/1105.3456} {arXiv:1105.3456 [gr-qc]} \BibitemShut
  {NoStop}%
\bibitem [{\citenamefont {{Ciufolini}}\ \emph {et~al.}(2016)\citenamefont
  {{Ciufolini}}, \citenamefont {{Paolozzi}}, \citenamefont {{Pavlis}},\ and\
  \citenamefont {{et al.}}}]{2016EPJC...76..120Cb}%
  \BibitemOpen
  \bibfield  {author} {\bibinfo {author} {\bibfnamefont {I.}~\bibnamefont
  {{Ciufolini}}}, \bibinfo {author} {\bibfnamefont {A.}~\bibnamefont
  {{Paolozzi}}}, \bibinfo {author} {\bibfnamefont {E.~C.}\ \bibnamefont
  {{Pavlis}}}, \ and\ \bibinfo {author} {\bibnamefont {{et al.}}},\ }\href
  {\doibase 10.1140/epjc/s10052-016-3961-8} {\bibfield  {journal} {\bibinfo
  {journal} {European Physical Journal C}\ }\textbf {\bibinfo {volume} {76}},\
  \bibinfo {eid} {120} (\bibinfo {year} {2016})},\ \Eprint
  {http://arxiv.org/abs/1603.09674} {arXiv:1603.09674 [gr-qc]} \BibitemShut
  {NoStop}%
\bibitem [{\citenamefont {{Lucchesi}}\ \emph {et~al.}(2017)\citenamefont
  {{Lucchesi}}, \citenamefont {{Magnafico}}, \citenamefont {{Peron}},\ and\
  \citenamefont {{et al.}}}]{2017mas..conf..131Lb}%
  \BibitemOpen
  \bibfield  {author} {\bibinfo {author} {\bibfnamefont {D.~M.}\ \bibnamefont
  {{Lucchesi}}}, \bibinfo {author} {\bibfnamefont {C.}~\bibnamefont
  {{Magnafico}}}, \bibinfo {author} {\bibfnamefont {R.}~\bibnamefont
  {{Peron}}}, \ and\ \bibinfo {author} {\bibnamefont {{et al.}}},\ }in\ \href
  {\doibase 10.1109/MetroAeroSpace.2017.7999552} {\emph {\bibinfo {booktitle}
  {{2017 IEEE International Workshop on Metrology for AeroSpace
  (MetroAeroSpace)}}}}\ (\bibinfo {year} {2017})\ p.\ \bibinfo {pages}
  {131}\BibitemShut {NoStop}%
\bibitem [{\citenamefont {{Schiff}}(1960{\natexlab{a}})}]{1960PhRvL...4..215S}%
  \BibitemOpen
  \bibfield  {author} {\bibinfo {author} {\bibfnamefont {L.~I.}\ \bibnamefont
  {{Schiff}}},\ }\href {\doibase 10.1103/PhysRevLett.4.215} {\bibfield
  {journal} {\bibinfo  {journal} {Physical Review Letters}\ }\textbf {\bibinfo
  {volume} {4}},\ \bibinfo {pages} {215} (\bibinfo {year}
  {1960}{\natexlab{a}})}\BibitemShut {NoStop}%
\bibitem [{\citenamefont {{Schiff}}(1960{\natexlab{b}})}]{1960PNAS...46..871S}%
  \BibitemOpen
  \bibfield  {author} {\bibinfo {author} {\bibfnamefont {L.~I.}\ \bibnamefont
  {{Schiff}}},\ }\href {\doibase 10.1073/pnas.46.6.871} {\bibfield  {journal}
  {\bibinfo  {journal} {Proceedings of the National Academy of Science}\
  }\textbf {\bibinfo {volume} {46}},\ \bibinfo {pages} {871} (\bibinfo {year}
  {1960}{\natexlab{b}})}\BibitemShut {NoStop}%
\bibitem [{\citenamefont {{Everitt}}\ \emph {et~al.}(2015)\citenamefont
  {{Everitt}}, \citenamefont {{Muhlfelder}}, \citenamefont {{DeBra}},\ and\
  \citenamefont {et~al.}}]{2015CQGra..32v4001Em}%
  \BibitemOpen
  \bibfield  {author} {\bibinfo {author} {\bibfnamefont {C.~W.~F.}\
  \bibnamefont {{Everitt}}}, \bibinfo {author} {\bibfnamefont {B.}~\bibnamefont
  {{Muhlfelder}}}, \bibinfo {author} {\bibfnamefont {D.~B.}\ \bibnamefont
  {{DeBra}}}, \ and\ \bibinfo {author} {\bibnamefont {et~al.}},\ }\href
  {\doibase 10.1088/0264-9381/32/22/224001} {\bibfield  {journal} {\bibinfo
  {journal} {Classical and Quantum Gravity}\ }\textbf {\bibinfo {volume}
  {32}},\ \bibinfo {eid} {224001} (\bibinfo {year} {2015})}\BibitemShut
  {NoStop}%
\bibitem [{\citenamefont {{Lucchesi}}(2007)}]{2007AdSpR..39..324L}%
  \BibitemOpen
  \bibfield  {author} {\bibinfo {author} {\bibfnamefont {D.~M.}\ \bibnamefont
  {{Lucchesi}}},\ }\href {\doibase 10.1016/j.asr.2006.10.012} {\bibfield
  {journal} {\bibinfo  {journal} {Adv. Space Res.}\ }\textbf {\bibinfo {volume}
  {39}},\ \bibinfo {pages} {324} (\bibinfo {year} {2007})}\BibitemShut
  {NoStop}%
\bibitem [{\citenamefont {{Lucchesi}}\ \emph {et~al.}(2015)\citenamefont
  {{Lucchesi}}, \citenamefont {{Anselmo}}, \citenamefont {{Bassan}},\ and\
  \citenamefont {{et al.}}}]{Lucchesietal2015b}%
  \BibitemOpen
  \bibfield  {author} {\bibinfo {author} {\bibfnamefont {D.}~\bibnamefont
  {{Lucchesi}}}, \bibinfo {author} {\bibfnamefont {L.}~\bibnamefont
  {{Anselmo}}}, \bibinfo {author} {\bibfnamefont {M.}~\bibnamefont {{Bassan}}},
  \ and\ \bibinfo {author} {\bibnamefont {{et al.}}},\ }\href {\doibase
  10.1088/0264-9381/32/15/155012} {\bibfield  {journal} {\bibinfo  {journal}
  {Class. Quantum Grav.}\ }\textbf {\bibinfo {volume} {32}},\ \bibinfo {pages}
  {155012} (\bibinfo {year} {2015})}\BibitemShut {NoStop}%
\bibitem [{\citenamefont {{Iorio}}(2003)}]{2003CeMDA..86..277I}%
  \BibitemOpen
  \bibfield  {author} {\bibinfo {author} {\bibfnamefont {L.}~\bibnamefont
  {{Iorio}}},\ }\href@noop {} {\bibfield  {journal} {\bibinfo  {journal}
  {Celestial Mechanics and Dynamical Astronomy}\ }\textbf {\bibinfo {volume}
  {86}},\ \bibinfo {pages} {277} (\bibinfo {year} {2003})},\ \Eprint
  {http://arxiv.org/abs/gr-qc/0203050} {gr-qc/0203050} \BibitemShut {NoStop}%
\bibitem [{\citenamefont {{Iorio}}(2005)}]{2005NewA...10..616I}%
  \BibitemOpen
  \bibfield  {author} {\bibinfo {author} {\bibfnamefont {L.}~\bibnamefont
  {{Iorio}}},\ }\href {\doibase 10.1016/j.newast.2005.02.006} {\bibfield
  {journal} {\bibinfo  {journal} {New Astron.}\ }\textbf {\bibinfo {volume}
  {10}},\ \bibinfo {pages} {616} (\bibinfo {year} {2005})},\ \Eprint
  {http://arxiv.org/abs/gr-qc/0502068} {gr-qc/0502068} \BibitemShut {NoStop}%
\bibitem [{\citenamefont {{Lucchesi}}(2005)}]{2005IJMPD..14.1989L}%
  \BibitemOpen
  \bibfield  {author} {\bibinfo {author} {\bibfnamefont {D.~M.}\ \bibnamefont
  {{Lucchesi}}},\ }\href {\doibase 10.1142/S0218271805008169} {\bibfield
  {journal} {\bibinfo  {journal} {International Journal of Modern Physics D}\
  }\textbf {\bibinfo {volume} {14}},\ \bibinfo {pages} {1989} (\bibinfo {year}
  {2005})}\BibitemShut {NoStop}%
\bibitem [{\citenamefont {{Iorio}}(2017)}]{2017EPJC...77...73I}%
  \BibitemOpen
  \bibfield  {author} {\bibinfo {author} {\bibfnamefont {L.}~\bibnamefont
  {{Iorio}}},\ }\href {\doibase 10.1140/epjc/s10052-017-4607-1} {\bibfield
  {journal} {\bibinfo  {journal} {European Physical Journal C}\ }\textbf
  {\bibinfo {volume} {77}},\ \bibinfo {eid} {73} (\bibinfo {year} {2017})},\
  \Eprint {http://arxiv.org/abs/1701.06474} {arXiv:1701.06474 [gr-qc]}
  \BibitemShut {NoStop}%
\bibitem [{\citenamefont {{Ciufolini}}\ \emph {et~al.}(2018)\citenamefont
  {{Ciufolini}}, \citenamefont {{Pavlis}}, \citenamefont {{Ries}},\ and\
  \citenamefont {{et al.}}}]{2018EPJC...78..880Cb}%
  \BibitemOpen
  \bibfield  {author} {\bibinfo {author} {\bibfnamefont {I.}~\bibnamefont
  {{Ciufolini}}}, \bibinfo {author} {\bibfnamefont {E.~C.}\ \bibnamefont
  {{Pavlis}}}, \bibinfo {author} {\bibfnamefont {J.}~\bibnamefont {{Ries}}}, \
  and\ \bibinfo {author} {\bibnamefont {{et al.}}},\ }\href {\doibase
  10.1140/epjc/s10052-018-6303-1} {\bibfield  {journal} {\bibinfo  {journal}
  {European Physical Journal C}\ }\textbf {\bibinfo {volume} {78}},\ \bibinfo
  {eid} {880} (\bibinfo {year} {2018})}\BibitemShut {NoStop}%
\bibitem [{\citenamefont {Lucchesi}\ \emph {et~al.}(2015)\citenamefont
  {Lucchesi}, \citenamefont {Peron}, \citenamefont {Visco},\ and\ \citenamefont
  {et~al.}}]{7180629b}%
  \BibitemOpen
  \bibfield  {author} {\bibinfo {author} {\bibfnamefont {D.~M.}\ \bibnamefont
  {Lucchesi}}, \bibinfo {author} {\bibfnamefont {R.}~\bibnamefont {Peron}},
  \bibinfo {author} {\bibfnamefont {M.}~\bibnamefont {Visco}}, \ and\ \bibinfo
  {author} {\bibnamefont {et~al.}},\ }in\ \href {\doibase
  10.1109/MetroAeroSpace.2015.7180629} {\emph {\bibinfo {booktitle} {{Metrology
  for Aerospace (MetroAeroSpace), 2015 IEEE}}}}\ (\bibinfo {year} {2015})\ pp.\
  \bibinfo {pages} {71--76}\BibitemShut {NoStop}%
\bibitem [{\citenamefont {{Lucchesi}}\ \emph {et~al.}(2019)\citenamefont
  {{Lucchesi}}, \citenamefont {{Anselmo}}, \citenamefont {{Bassan}},
  \citenamefont {{Magnafico}}, \citenamefont {{Pardini}}, \citenamefont
  {{Peron}}, \citenamefont {{Pucacco}},\ and\ \citenamefont
  {{Visco}}}]{2019Univ....5..141L}%
  \BibitemOpen
  \bibfield  {author} {\bibinfo {author} {\bibfnamefont {D.~M.}\ \bibnamefont
  {{Lucchesi}}}, \bibinfo {author} {\bibfnamefont {L.}~\bibnamefont
  {{Anselmo}}}, \bibinfo {author} {\bibfnamefont {M.}~\bibnamefont {{Bassan}}},
  \bibinfo {author} {\bibfnamefont {C.}~\bibnamefont {{Magnafico}}}, \bibinfo
  {author} {\bibfnamefont {C.}~\bibnamefont {{Pardini}}}, \bibinfo {author}
  {\bibfnamefont {R.}~\bibnamefont {{Peron}}}, \bibinfo {author} {\bibfnamefont
  {G.}~\bibnamefont {{Pucacco}}}, \ and\ \bibinfo {author} {\bibfnamefont
  {M.}~\bibnamefont {{Visco}}},\ }\href {\doibase 10.3390/universe5060141}
  {\bibfield  {journal} {\bibinfo  {journal} {Universe}\ }\textbf {\bibinfo
  {volume} {5}},\ \bibinfo {pages} {141} (\bibinfo {year} {2019})}\BibitemShut
  {NoStop}%
\bibitem [{\citenamefont {Pucacco}\ \emph {et~al.}(2017)\citenamefont
  {Pucacco}, \citenamefont {Lucchesi}, \citenamefont {Anselmo},\ and\
  \citenamefont {et~al.}}]{LucchesiEGU2017p2b}%
  \BibitemOpen
  \bibfield  {author} {\bibinfo {author} {\bibfnamefont {G.}~\bibnamefont
  {Pucacco}}, \bibinfo {author} {\bibfnamefont {D.~M.}\ \bibnamefont
  {Lucchesi}}, \bibinfo {author} {\bibfnamefont {L.}~\bibnamefont {Anselmo}}, \
  and\ \bibinfo {author} {\bibnamefont {et~al.}},\ }in\ \href@noop {} {\emph
  {\bibinfo {booktitle} {{EGU Conference}}}},\ \bibinfo {series and number}
  {{Geophysical Research Abstracts, Vol. 19, EGU2017-13554}}\ (\bibinfo {year}
  {2017})\BibitemShut {NoStop}%
\bibitem [{\citenamefont {{Pucacco}}\ and\ \citenamefont
  {{Lucchesi}}(2018)}]{2018CeMDA.130...66P}%
  \BibitemOpen
  \bibfield  {author} {\bibinfo {author} {\bibfnamefont {G.}~\bibnamefont
  {{Pucacco}}}\ and\ \bibinfo {author} {\bibfnamefont {D.~M.}\ \bibnamefont
  {{Lucchesi}}},\ }\href {\doibase 10.1007/s10569-018-9861-5} {\bibfield
  {journal} {\bibinfo  {journal} {Celestial Mechanics and Dynamical Astronomy}\
  }\textbf {\bibinfo {volume} {130}},\ \bibinfo {eid} {66} (\bibinfo {year}
  {2018})}\BibitemShut {NoStop}%
\bibitem [{\citenamefont {Pucacco}\ \emph {et~al.}(2019)\citenamefont
  {Pucacco}, \citenamefont {Lucchesi}, \citenamefont {Anselmo}, \citenamefont
  {Bassan}, \citenamefont {Magnafico}, \citenamefont {Pardini}, \citenamefont
  {Peron}, \citenamefont {Stanga},\ and\ \citenamefont
  {Visco}}]{LucchesiEGU2019p2}%
  \BibitemOpen
  \bibfield  {author} {\bibinfo {author} {\bibfnamefont {G.}~\bibnamefont
  {Pucacco}}, \bibinfo {author} {\bibfnamefont {D.~M.}\ \bibnamefont
  {Lucchesi}}, \bibinfo {author} {\bibfnamefont {L.}~\bibnamefont {Anselmo}},
  \bibinfo {author} {\bibfnamefont {M.}~\bibnamefont {Bassan}}, \bibinfo
  {author} {\bibfnamefont {C.}~\bibnamefont {Magnafico}}, \bibinfo {author}
  {\bibfnamefont {C.}~\bibnamefont {Pardini}}, \bibinfo {author} {\bibfnamefont
  {R.}~\bibnamefont {Peron}}, \bibinfo {author} {\bibfnamefont
  {R.}~\bibnamefont {Stanga}}, \ and\ \bibinfo {author} {\bibfnamefont
  {M.}~\bibnamefont {Visco}},\ }in\ \href@noop {} {\emph {\bibinfo {booktitle}
  {{EGU Conference}}}},\ \bibinfo {series and number} {{Geophysical Research
  Abstracts, Vol. 21, EGU2019-10721}}\ (\bibinfo {year} {2019})\BibitemShut
  {NoStop}%
\bibitem [{\citenamefont {{Visco}}\ and\ \citenamefont
  {{Lucchesi}}(2016)}]{2016AdSpR..57.1928V}%
  \BibitemOpen
  \bibfield  {author} {\bibinfo {author} {\bibfnamefont {M.}~\bibnamefont
  {{Visco}}}\ and\ \bibinfo {author} {\bibfnamefont {D.~M.}\ \bibnamefont
  {{Lucchesi}}},\ }\href {\doibase 10.1016/j.asr.2016.02.006} {\bibfield
  {journal} {\bibinfo  {journal} {Advances in Space Research}\ }\textbf
  {\bibinfo {volume} {57}},\ \bibinfo {pages} {1928} (\bibinfo {year}
  {2016})}\BibitemShut {NoStop}%
\bibitem [{\citenamefont {{Pardini}}\ \emph {et~al.}(2017)\citenamefont
  {{Pardini}}, \citenamefont {{Anselmo}}, \citenamefont {{Lucchesi}},\ and\
  \citenamefont {{Peron}}}]{2017AcAau.140..469P}%
  \BibitemOpen
  \bibfield  {author} {\bibinfo {author} {\bibfnamefont {C.}~\bibnamefont
  {{Pardini}}}, \bibinfo {author} {\bibfnamefont {L.}~\bibnamefont
  {{Anselmo}}}, \bibinfo {author} {\bibfnamefont {D.~M.}\ \bibnamefont
  {{Lucchesi}}}, \ and\ \bibinfo {author} {\bibfnamefont {R.}~\bibnamefont
  {{Peron}}},\ }\href {\doibase 10.1016/j.actaastro.2017.09.012} {\bibfield
  {journal} {\bibinfo  {journal} {Acta Astronautica}\ }\textbf {\bibinfo
  {volume} {140}},\ \bibinfo {pages} {469} (\bibinfo {year}
  {2017})}\BibitemShut {NoStop}%
\bibitem [{\citenamefont {{Visco}}\ and\ \citenamefont
  {{Lucchesi}}(2018)}]{2018PhRvD..98d4034V}%
  \BibitemOpen
  \bibfield  {author} {\bibinfo {author} {\bibfnamefont {M.}~\bibnamefont
  {{Visco}}}\ and\ \bibinfo {author} {\bibfnamefont {D.~M.}\ \bibnamefont
  {{Lucchesi}}},\ }\href {\doibase 10.1103/PhysRevD.98.044034} {\bibfield
  {journal} {\bibinfo  {journal} {\prd}\ }\textbf {\bibinfo {volume} {98}},\
  \bibinfo {eid} {044034} (\bibinfo {year} {2018})}\BibitemShut {NoStop}%
\bibitem [{\citenamefont {Pardini}\ \emph {et~al.}(2019)\citenamefont
  {Pardini}, \citenamefont {Anselmo}, \citenamefont {Lucchesi},\ and\
  \citenamefont {et~al.}}]{PardiniEGU2019p2b}%
  \BibitemOpen
  \bibfield  {author} {\bibinfo {author} {\bibfnamefont {C.}~\bibnamefont
  {Pardini}}, \bibinfo {author} {\bibfnamefont {L.}~\bibnamefont {Anselmo}},
  \bibinfo {author} {\bibfnamefont {D.~M.}\ \bibnamefont {Lucchesi}}, \ and\
  \bibinfo {author} {\bibnamefont {et~al.}},\ }in\ \href@noop {} {\emph
  {\bibinfo {booktitle} {{EGU Conference}}}},\ \bibinfo {series and number}
  {{Geophysical Research Abstracts, Vol. 21, EGU2019-16897}}\ (\bibinfo {year}
  {2019})\BibitemShut {NoStop}%
\bibitem [{\citenamefont {{Kozai}}(1959)}]{1959AJ.....64..367K}%
  \BibitemOpen
  \bibfield  {author} {\bibinfo {author} {\bibfnamefont {Y.}~\bibnamefont
  {{Kozai}}},\ }\href {\doibase 10.1086/107957} {\bibfield  {journal} {\bibinfo
   {journal} {Astron. J.}\ }\textbf {\bibinfo {volume} {64}},\ \bibinfo {pages}
  {367} (\bibinfo {year} {1959})}\BibitemShut {NoStop}%
\bibitem [{\citenamefont {{Kaula}}(1966)}]{1966tsga.book.....K}%
  \BibitemOpen
  \bibfield  {author} {\bibinfo {author} {\bibfnamefont {W.~M.}\ \bibnamefont
  {{Kaula}}},\ }\href@noop {} {\emph {\bibinfo {title} {{Theory of satellite
  geodesy. Applications of satellites to geodesy}}}}\ (\bibinfo  {publisher}
  {Blaisdell},\ \bibinfo {address} {Waltham, Mass.},\ \bibinfo {year}
  {1966})\BibitemShut {NoStop}%
\bibitem [{Note1()}]{Note1}%
  \BibitemOpen
  \bibinfo {note} {The order-of-magnitude of this \protect \textit {classical}
  precession is about \(+126\) deg./yr for LAGEOS, \(-231\) deg./yr for LAGEOS
  II, and about \(-624\) deg./yr in the case of LARES.}\BibitemShut {Stop}%
\bibitem [{\citenamefont {{Cheng}}\ \emph {et~al.}(1997)\citenamefont
  {{Cheng}}, \citenamefont {{Shum}},\ and\ \citenamefont
  {{Tapley}}}]{1997JGR...10222377C}%
  \BibitemOpen
  \bibfield  {author} {\bibinfo {author} {\bibfnamefont {M.~K.}\ \bibnamefont
  {{Cheng}}}, \bibinfo {author} {\bibfnamefont {C.~K.}\ \bibnamefont {{Shum}}},
  \ and\ \bibinfo {author} {\bibfnamefont {B.~D.}\ \bibnamefont {{Tapley}}},\
  }\href {\doibase 10.1029/97JB01740} {\bibfield  {journal} {\bibinfo
  {journal} {J. Geophys. Res.}\ }\textbf {\bibinfo {volume} {102}},\ \bibinfo
  {pages} {22377} (\bibinfo {year} {1997})}\BibitemShut {NoStop}%
\bibitem [{\citenamefont {{Cox}}\ and\ \citenamefont
  {{Chao}}(2002)}]{2002Sci...297..831C}%
  \BibitemOpen
  \bibfield  {author} {\bibinfo {author} {\bibfnamefont {C.~M.}\ \bibnamefont
  {{Cox}}}\ and\ \bibinfo {author} {\bibfnamefont {B.~F.}\ \bibnamefont
  {{Chao}}},\ }\href {\doibase 10.1126/science.1072188} {\bibfield  {journal}
  {\bibinfo  {journal} {Science}\ }\textbf {\bibinfo {volume} {297}},\ \bibinfo
  {pages} {831} (\bibinfo {year} {2002})}\BibitemShut {NoStop}%
\bibitem [{\citenamefont {Cheng}\ \emph {et~al.}(2013)\citenamefont {Cheng},
  \citenamefont {Tapley},\ and\ \citenamefont {Ries}}]{JGRB:JGRB50058}%
  \BibitemOpen
  \bibfield  {author} {\bibinfo {author} {\bibfnamefont {M.}~\bibnamefont
  {Cheng}}, \bibinfo {author} {\bibfnamefont {B.~D.}\ \bibnamefont {Tapley}}, \
  and\ \bibinfo {author} {\bibfnamefont {J.~C.}\ \bibnamefont {Ries}},\ }\href
  {\doibase 10.1002/jgrb.50058} {\bibfield  {journal} {\bibinfo  {journal}
  {Journal of Geophysical Research: Solid Earth}\ }\textbf {\bibinfo {volume}
  {118}},\ \bibinfo {pages} {740} (\bibinfo {year} {2013})}\BibitemShut
  {NoStop}%
\bibitem [{\citenamefont {{Cheng}}\ and\ \citenamefont
  {{Ries}}(2018)}]{2018GeoJI.212.1218C}%
  \BibitemOpen
  \bibfield  {author} {\bibinfo {author} {\bibfnamefont {M.}~\bibnamefont
  {{Cheng}}}\ and\ \bibinfo {author} {\bibfnamefont {J.~C.}\ \bibnamefont
  {{Ries}}},\ }\href {\doibase 10.1093/gji/ggx483} {\bibfield  {journal}
  {\bibinfo  {journal} {Geophysical Journal International}\ }\textbf {\bibinfo
  {volume} {212}},\ \bibinfo {pages} {1218} (\bibinfo {year}
  {2018})}\BibitemShut {NoStop}%
\bibitem [{Note2()}]{Note2}%
  \BibitemOpen
  \bibinfo {note} {In \cite {2019Univ....5..141L}, we fitted linearly the
  quadrupole coefficient obtained from GRACE (Gravity Recovery And Climate
  Experiment) monthly solutions, and we used this fitted value in the data
  reduction of the satellites orbit. Furthermore, in that paper, and for the
  GGM05S model, with regard to the LT effect measurement, we have also analyzed
  and compared the error related to the knowledge of the octupole coefficient
  with respect to the hexapole one.}\BibitemShut {Stop}%
\bibitem [{\citenamefont {{Reigber}}\ \emph {et~al.}(2002)\citenamefont
  {{Reigber}}, \citenamefont {{L{\"u}hr}},\ and\ \citenamefont
  {{Schwintzer}}}]{2002AdSpR..30..129R}%
  \BibitemOpen
  \bibfield  {author} {\bibinfo {author} {\bibfnamefont {C.}~\bibnamefont
  {{Reigber}}}, \bibinfo {author} {\bibfnamefont {H.}~\bibnamefont
  {{L{\"u}hr}}}, \ and\ \bibinfo {author} {\bibfnamefont {P.}~\bibnamefont
  {{Schwintzer}}},\ }\href {\doibase 10.1016/S0273-1177(02)00276-4} {\bibfield
  {journal} {\bibinfo  {journal} {Adv. Space Res.}\ }\textbf {\bibinfo {volume}
  {30}},\ \bibinfo {pages} {129} (\bibinfo {year} {2002})}\BibitemShut
  {NoStop}%
\bibitem [{\citenamefont {{Reigber}}\ \emph {et~al.}(2003)\citenamefont
  {{Reigber}}, \citenamefont {{Schwintzer}}, \citenamefont {{Neumayer}},\ and\
  \citenamefont {{et. al.}}}]{2003AdSpR..31.1883Rb}%
  \BibitemOpen
  \bibfield  {author} {\bibinfo {author} {\bibfnamefont {C.}~\bibnamefont
  {{Reigber}}}, \bibinfo {author} {\bibfnamefont {P.}~\bibnamefont
  {{Schwintzer}}}, \bibinfo {author} {\bibfnamefont {K.-H.}\ \bibnamefont
  {{Neumayer}}}, \ and\ \bibinfo {author} {\bibnamefont {{et. al.}}},\ }\href
  {\doibase 10.1016/S0273-1177(03)00162-5} {\bibfield  {journal} {\bibinfo
  {journal} {Adv. Space Res.}\ }\textbf {\bibinfo {volume} {31}},\ \bibinfo
  {pages} {1883} (\bibinfo {year} {2003})}\BibitemShut {NoStop}%
\bibitem [{\citenamefont {{Reigber}}\ \emph {et~al.}(2005)\citenamefont
  {{Reigber}}, \citenamefont {{Schmidt}}, \citenamefont {{Flechtner}},\ and\
  \citenamefont {{et al.}}}]{2005JGeo...39....1Rb}%
  \BibitemOpen
  \bibfield  {author} {\bibinfo {author} {\bibfnamefont {C.}~\bibnamefont
  {{Reigber}}}, \bibinfo {author} {\bibfnamefont {R.}~\bibnamefont
  {{Schmidt}}}, \bibinfo {author} {\bibfnamefont {F.}~\bibnamefont
  {{Flechtner}}}, \ and\ \bibinfo {author} {\bibnamefont {{et al.}}},\ }\href
  {\doibase 10.1016/j.jog.2004.07.001} {\bibfield  {journal} {\bibinfo
  {journal} {J. Geodyn.}\ }\textbf {\bibinfo {volume} {39}},\ \bibinfo {pages}
  {1} (\bibinfo {year} {2005})}\BibitemShut {NoStop}%
\bibitem [{\citenamefont {{Tapley}}\ and\ \citenamefont
  {{Reigber}}(2001)}]{2001AGUFM.G41C..02T}%
  \BibitemOpen
  \bibfield  {author} {\bibinfo {author} {\bibfnamefont {B.~D.}\ \bibnamefont
  {{Tapley}}}\ and\ \bibinfo {author} {\bibfnamefont {C.}~\bibnamefont
  {{Reigber}}},\ }\href@noop {} {\bibfield  {journal} {\bibinfo  {journal} {AGU
  Fall Meeting Abstracts}\ ,\ \bibinfo {pages} {C2}} (\bibinfo {year}
  {2001})}\BibitemShut {NoStop}%
\bibitem [{\citenamefont {{Lemoine}}\ \emph {et~al.}(1998)\citenamefont
  {{Lemoine}}, \citenamefont {{Kenyon}}, \citenamefont {{Factor}},\ and\
  \citenamefont {{et al.}}}]{1998Lemoineb}%
  \BibitemOpen
  \bibfield  {author} {\bibinfo {author} {\bibfnamefont {F.~G.}\ \bibnamefont
  {{Lemoine}}}, \bibinfo {author} {\bibfnamefont {S.}~\bibnamefont {{Kenyon}}},
  \bibinfo {author} {\bibfnamefont {J.~K.}\ \bibnamefont {{Factor}}}, \ and\
  \bibinfo {author} {\bibnamefont {{et al.}}},\ }\href@noop {} {\emph {\bibinfo
  {title} {{The Development of the Joint NASA GSFC and the National Imagery and
  Mapping Agency (NIMA) Geopotential Model EGM96}}}},\ \bibinfo {type}
  {Technical Paper}\ \bibinfo {number} {206861}\ (\bibinfo  {institution}
  {NASA},\ \bibinfo {year} {1998})\BibitemShut {NoStop}%
\bibitem [{Note3()}]{Note3}%
  \BibitemOpen
  \bibinfo {note} {See the International Centre for Global Earth Models
  (ICGEM): Gravity Field Solutions for dedicated Time Periods: Release 05,
  http://icgem.gfz-potsdam.de/series (2018).}\BibitemShut {Stop}%
\bibitem [{Note4()}]{Note4}%
  \BibitemOpen
  \bibinfo {note} {This was also extended up to degree \(\ell =30\), but with
  no appreciable difference in the results. In the case of the quadrupole
  coefficient we also considered a more complete non-linear fit to the time
  behaviour outlined by GRACE, with the inclusion of two periodic terms, one
  with a yearly frequency and one at twice this frequency. However, the final
  results have not shown a noticeable difference with respect to those obtained
  with a simpler linear fit.}\BibitemShut {Stop}%
\bibitem [{\citenamefont {{Petit}}\ and\ \citenamefont
  {{Luzum}}(2010)}]{2010IERS-Conv-2010}%
  \BibitemOpen
  \bibfield  {author} {\bibinfo {author} {\bibfnamefont {G.}~\bibnamefont
  {{Petit}}}\ and\ \bibinfo {author} {\bibfnamefont {B.}~\bibnamefont
  {{Luzum}}},\ }\href@noop {} {\emph {\bibinfo {title} {{IERS Conventions
  (2010)}}}},\ \bibinfo {type} {IERS Technical Note}\ \bibinfo {number} {36}\
  (\bibinfo  {institution} {IERS},\ \bibinfo {address} {Frankfurt am Main:
  Verlag des Bundesamts f{\"u}r Kartographie und Geod{\"a}sie},\ \bibinfo
  {year} {2010})\BibitemShut {NoStop}%
\bibitem [{Note5()}]{Note5}%
  \BibitemOpen
  \bibinfo {note} {Usually, these trends were those suggested by IERS
  Conventions~\cite {2010IERS-Conv-2010}, but were not always compatible with
  the results from GRACE monthly solutions.}\BibitemShut {Stop}%
\bibitem [{\citenamefont {{Chen}}\ \emph {et~al.}(2013)\citenamefont {{Chen}},
  \citenamefont {{Wilson}}, \citenamefont {{Ries}},\ and\ \citenamefont
  {{Tapley}}}]{2013GeoRL..40.2625C}%
  \BibitemOpen
  \bibfield  {author} {\bibinfo {author} {\bibfnamefont {J.~L.}\ \bibnamefont
  {{Chen}}}, \bibinfo {author} {\bibfnamefont {C.~R.}\ \bibnamefont
  {{Wilson}}}, \bibinfo {author} {\bibfnamefont {J.~C.}\ \bibnamefont
  {{Ries}}}, \ and\ \bibinfo {author} {\bibfnamefont {B.~D.}\ \bibnamefont
  {{Tapley}}},\ }\href {\doibase 10.1002/grl.50552} {\bibfield  {journal}
  {\bibinfo  {journal} {Geophys. Res. Lett.}\ }\textbf {\bibinfo {volume}
  {40}},\ \bibinfo {pages} {2625} (\bibinfo {year} {2013})}\BibitemShut
  {NoStop}%
\bibitem [{Note6()}]{Note6}%
  \BibitemOpen
  \bibinfo {note} {It is important to stress that the static models provide, in
  general, good measurements for the coefficients of the Earth's gravity field
  on the time span of GRACE data over which they have been effectively
  computed, i.e. they represent averages values on this time interval. However,
  on different subintervals of the entire period of GRACE data, as in the case
  of our analysis, that starts after the launch of the LARES satellites, these
  average values may be quite different with respect to their effective time
  behaviour provided by GRACE monthly solutions.}\BibitemShut {Stop}%
\bibitem [{\citenamefont {{Pavlis}}\ and\ \citenamefont {{et
  al.}}(1998)}]{1998pavlis}%
  \BibitemOpen
  \bibfield  {author} {\bibinfo {author} {\bibfnamefont {D.~E.}\ \bibnamefont
  {{Pavlis}}}\ and\ \bibinfo {author} {\bibnamefont {{et al.}}},\ }\href@noop
  {} {\emph {\bibinfo {title} {{GEODYN II Operations Manual}}}},\ \bibinfo
  {organization} {NASA GSFC} (\bibinfo {year} {1998})\BibitemShut {NoStop}%
\bibitem [{\citenamefont {{Putney}}\ \emph {et~al.}(1990)\citenamefont
  {{Putney}}, \citenamefont {{Kolenkiewicz}}, \citenamefont {{Smith}},
  \citenamefont {{Dunn}},\ and\ \citenamefont
  {{Torrence}}}]{1990AdSpR..10..197P}%
  \BibitemOpen
  \bibfield  {author} {\bibinfo {author} {\bibfnamefont {B.}~\bibnamefont
  {{Putney}}}, \bibinfo {author} {\bibfnamefont {R.}~\bibnamefont
  {{Kolenkiewicz}}}, \bibinfo {author} {\bibfnamefont {D.}~\bibnamefont
  {{Smith}}}, \bibinfo {author} {\bibfnamefont {P.}~\bibnamefont {{Dunn}}}, \
  and\ \bibinfo {author} {\bibfnamefont {M.~H.}\ \bibnamefont {{Torrence}}},\
  }\href {\doibase 10.1016/0273-1177(90)90350-9} {\bibfield  {journal}
  {\bibinfo  {journal} {Adv. Space Res.}\ }\textbf {\bibinfo {volume} {10}},\
  \bibinfo {pages} {197} (\bibinfo {year} {1990})}\BibitemShut {NoStop}%
\bibitem [{\citenamefont {{Sinclair}}(1997)}]{Sinclair1997}%
  \BibitemOpen
  \bibfield  {author} {\bibinfo {author} {\bibfnamefont {A.~T.}\ \bibnamefont
  {{Sinclair}}},\ }\href
  {http://ilrs.gsfc.nasa.gov/products_formats_procedures/normal_point/np_algo.html}
  {\enquote {\bibinfo {title} {{Data Screening and Normal Point Formation ---
  Re--Statement of Herstmonceux Normal Point Recommendation}},}\ } (\bibinfo
  {year} {1997})\BibitemShut {NoStop}%
\bibitem [{\citenamefont {{Huang}}\ \emph {et~al.}(1990)\citenamefont
  {{Huang}}, \citenamefont {{Ries}}, \citenamefont {{Tapley}},\ and\
  \citenamefont {{Watkins}}}]{1990CeMDA..48..167H}%
  \BibitemOpen
  \bibfield  {author} {\bibinfo {author} {\bibfnamefont {C.}~\bibnamefont
  {{Huang}}}, \bibinfo {author} {\bibfnamefont {J.~C.}\ \bibnamefont {{Ries}}},
  \bibinfo {author} {\bibfnamefont {B.~D.}\ \bibnamefont {{Tapley}}}, \ and\
  \bibinfo {author} {\bibfnamefont {M.~M.}\ \bibnamefont {{Watkins}}},\ }\href
  {\doibase 10.1007/BF00049512} {\bibfield  {journal} {\bibinfo  {journal}
  {Celest. Mech. Dyn. Astron.}\ }\textbf {\bibinfo {volume} {48}},\ \bibinfo
  {pages} {167} (\bibinfo {year} {1990})}\BibitemShut {NoStop}%
\bibitem [{\citenamefont {{Lucchesi}}(2002)}]{2002P&SS...50.1067L}%
  \BibitemOpen
  \bibfield  {author} {\bibinfo {author} {\bibfnamefont {D.~M.}\ \bibnamefont
  {{Lucchesi}}},\ }\href {\doibase 10.1016/S0032-0633(02)00052-1} {\bibfield
  {journal} {\bibinfo  {journal} {Plan. Space Sci.}\ }\textbf {\bibinfo
  {volume} {50}},\ \bibinfo {pages} {1067} (\bibinfo {year} {2002})},\ \bibinfo
  {note} {presented at OCA/CERGA Observatory, Grasse, France, December 18,
  2001}\BibitemShut {NoStop}%
\bibitem [{Note7()}]{Note7}%
  \BibitemOpen
  \bibinfo {note} {This exclusion is motivated, due to the complexity of the
  effects and their dependence on the spin vector evolution of the satellite,
  by the fact that the routines included in GEODYN for their modelling are not
  updated. Moreover they were valid only in the fast rotation regime of the
  satellites, which is not longer applicable for the two LAGEOS satellites
  \cite {2013AdSpR..52.1332K,2018PhRvD..98d4034V}. For the same reason we had
  to exclude the pericenter from our analysis. Our ongoing effort to improve
  the model for the thermal thrust perturbations~\cite
  {2002P&SS...50.1067L,2016AdSpR..57.1928V,2018PhRvD..98d4034V,2019Univ....5..141L}
  will hopefully allow us to include, in forthcoming measurements, also this
  element in the analysis, especially for LAGEOS II.}\BibitemShut {Stop}%
\bibitem [{\citenamefont {{Tapley}}\ \emph {et~al.}(2013)\citenamefont
  {{Tapley}}, \citenamefont {{Flechtner}}, \citenamefont {{Bettadpur}},\ and\
  \citenamefont {{Watkins}}}]{Tapley2013}%
  \BibitemOpen
  \bibfield  {author} {\bibinfo {author} {\bibfnamefont {B.~D.}\ \bibnamefont
  {{Tapley}}}, \bibinfo {author} {\bibfnamefont {F.}~\bibnamefont
  {{Flechtner}}}, \bibinfo {author} {\bibfnamefont {S.~V.}\ \bibnamefont
  {{Bettadpur}}}, \ and\ \bibinfo {author} {\bibfnamefont {M.~M.}\ \bibnamefont
  {{Watkins}}},\ }\href@noop {} {\bibfield  {journal} {\bibinfo  {journal} {Eos
  Trans. Fall Meet. Suppl. Abstract G22A-01}\ } (\bibinfo {year}
  {2013})}\BibitemShut {NoStop}%
\bibitem [{Note8()}]{Note8}%
  \BibitemOpen
  \bibinfo {note} {This is the solution currently used by the International
  Laser Ranging Service (ILRS) to realize the International Terrestrial
  Reference Frame (ITRF2014) \cite {2016JGRB..121.6109A}, i.e. the practical
  realization of the International Terrestrial Reference System (ITRS)~\cite
  {2010IERS-Conv-2010}.}\BibitemShut {Stop}%
\bibitem [{\citenamefont {{Akyilmaz}}\ \emph {et~al.}(2016)\citenamefont
  {{Akyilmaz}}, \citenamefont {{Ustun}}, \citenamefont {{Aydin}},\ and\
  \citenamefont {{et al.}}}]{ITU-GRACE16b}%
  \BibitemOpen
  \bibfield  {author} {\bibinfo {author} {\bibfnamefont {O.}~\bibnamefont
  {{Akyilmaz}}}, \bibinfo {author} {\bibfnamefont {A.}~\bibnamefont {{Ustun}}},
  \bibinfo {author} {\bibfnamefont {C.}~\bibnamefont {{Aydin}}}, \ and\
  \bibinfo {author} {\bibnamefont {{et al.}}},\ }\href {\doibase
  10.5880/icgem.2016.006} {\bibfield  {journal} {\bibinfo  {journal} {GFZ Data
  Services}\ } (\bibinfo {year} {2016}),\ 10.5880/icgem.2016.006}\BibitemShut
  {NoStop}%
\bibitem [{\citenamefont {{Chen}}\ \emph {et~al.}(2017)\citenamefont {{Chen}},
  \citenamefont {{Shen}}, \citenamefont {{Chen}},\ and\ \citenamefont
  {{Zhang}}}]{Tongji-Grace02s}%
  \BibitemOpen
  \bibfield  {author} {\bibinfo {author} {\bibfnamefont {Q.}~\bibnamefont
  {{Chen}}}, \bibinfo {author} {\bibfnamefont {Y.}~\bibnamefont {{Shen}}},
  \bibinfo {author} {\bibfnamefont {W.}~\bibnamefont {{Chen}}}, \ and\ \bibinfo
  {author} {\bibfnamefont {X.}~\bibnamefont {{Zhang}}},\ }\href {\doibase
  10.5880/icgem.2017.002} {\bibfield  {journal} {\bibinfo  {journal} {GFZ Data
  Services}\ } (\bibinfo {year} {2017}),\ 10.5880/icgem.2017.002}\BibitemShut
  {NoStop}%
\bibitem [{\citenamefont {{Chen}}\ \emph {et~al.}(2018)\citenamefont {{Chen}},
  \citenamefont {{Shen}}, \citenamefont {{Francis}}, \citenamefont {{Chen}},
  \citenamefont {{Zhang}},\ and\ \citenamefont {{Hsu}}}]{2018JGRB..123.6111C}%
  \BibitemOpen
  \bibfield  {author} {\bibinfo {author} {\bibfnamefont {Q.}~\bibnamefont
  {{Chen}}}, \bibinfo {author} {\bibfnamefont {Y.}~\bibnamefont {{Shen}}},
  \bibinfo {author} {\bibfnamefont {O.}~\bibnamefont {{Francis}}}, \bibinfo
  {author} {\bibfnamefont {W.}~\bibnamefont {{Chen}}}, \bibinfo {author}
  {\bibfnamefont {X.}~\bibnamefont {{Zhang}}}, \ and\ \bibinfo {author}
  {\bibfnamefont {H.}~\bibnamefont {{Hsu}}},\ }\href {\doibase
  10.1029/2018JB015641} {\bibfield  {journal} {\bibinfo  {journal} {Journal of
  Geophysical Research (Solid Earth)}\ }\textbf {\bibinfo {volume} {123}},\
  \bibinfo {pages} {6111} (\bibinfo {year} {2018})}\BibitemShut {NoStop}%
\bibitem [{\citenamefont {{Colombo}}(1984)}]{1984aot..book.....C}%
  \BibitemOpen
  \bibfield  {author} {\bibinfo {author} {\bibfnamefont {O.~L.}\ \bibnamefont
  {{Colombo}}},\ }\href@noop {} {\emph {\bibinfo {title} {{NASA Tech.~Memo.,
  NASA TM-86180, 9+173 pp.}}}}\ (\bibinfo {year} {1984})\BibitemShut {NoStop}%
\bibitem [{\citenamefont {{Ray}}(1999)}]{1999Ray}%
  \BibitemOpen
  \bibfield  {author} {\bibinfo {author} {\bibfnamefont {R.~D.}\ \bibnamefont
  {{Ray}}},\ }\href@noop {} {\emph {\bibinfo {title} {{A Global Ocean Tide
  Model From TOPEX/POSEIDON Altimetry: GOT99.2}}}},\ \bibinfo {type} {Technical
  Paper}\ \bibinfo {number} {NASA/TM-1999-209478}\ (\bibinfo  {institution}
  {Goddard Space Flight Center, Greenbelt, Maryland},\ \bibinfo {year}
  {1999})\BibitemShut {NoStop}%
\bibitem [{ilr(2018)}]{ilrs2}%
  \BibitemOpen
  \href@noop {} {\emph {\bibinfo {title} {{ILRS recommendations: Data
  Corrections}}}},\ \bibinfo {type}
  {https://ilrs.cddis.eosdis.nasa.gov/network/site\_information/}\ (\bibinfo
  {year} {2018})\BibitemShut {NoStop}%
\bibitem [{Note9()}]{Note9}%
  \BibitemOpen
  \bibinfo {note} {We have estimated the LT effect from the orbit of LAGEOS,
  LAGEOS II and LARES applying several different strategies. Some of these
  correspond to new methods with respect to those reported in the published
  literature on the LT effect measurements. We cannot present all these methods
  in this Letter, but we can disclose that the final results obtained for \(\mu
  \) are all consistent, with a precision at the level of a fraction of percent
  for the cumulative residuals.}\BibitemShut {Stop}%
\bibitem [{\citenamefont {{Ciufolini}}(1996)}]{1996NCimA.109.1709C}%
  \BibitemOpen
  \bibfield  {author} {\bibinfo {author} {\bibfnamefont {I.}~\bibnamefont
  {{Ciufolini}}},\ }\href {\doibase 10.1007/BF02773551} {\bibfield  {journal}
  {\bibinfo  {journal} {Nuovo Cim. A}\ }\textbf {\bibinfo {volume} {109}},\
  \bibinfo {pages} {1709} (\bibinfo {year} {1996})}\BibitemShut {NoStop}%
\bibitem [{Note10()}]{Note10}%
  \BibitemOpen
  \bibinfo {note} {In this system of Eqs. we are neglecting the contributions
  from the mismodeling of the higher harmonics. Anyway, their contribution is
  explicitly considered in the evaluation of the systematic errors, i.e. in the
  overall error budget of the measurement.}\BibitemShut {Stop}%
\bibitem [{\citenamefont {{Lucchesi}}\ and\ \citenamefont
  {{Balmino}}(2006)}]{2006P&SS...54..581L}%
  \BibitemOpen
  \bibfield  {author} {\bibinfo {author} {\bibfnamefont {D.~M.}\ \bibnamefont
  {{Lucchesi}}}\ and\ \bibinfo {author} {\bibfnamefont {G.}~\bibnamefont
  {{Balmino}}},\ }\href {\doibase 10.1016/j.pss.2006.03.001} {\bibfield
  {journal} {\bibinfo  {journal} {Plan. Space Sci.}\ }\textbf {\bibinfo
  {volume} {54}},\ \bibinfo {pages} {581} (\bibinfo {year} {2006})}\BibitemShut
  {NoStop}%
\bibitem [{Note11()}]{Note11}%
  \BibitemOpen
  \bibinfo {note} {The orbital parameters are known with a very small relative
  uncertainty, such that the \(K\) coefficients can be considered, for our
  purposes, as error-free.}\BibitemShut {Stop}%
\bibitem [{Note12()}]{Note12}%
  \BibitemOpen
  \bibinfo {note} {In this Letter we directly provide the result for \(\mu \)
  and not, as done in previous measurements of the LT effect, for the combined
  rate of the RAAN of the three satellites. In our analyses, this combination
  corresponds to a precession of 50.17 mas/yr.}\BibitemShut {Stop}%
\end{thebibliography}%

\end{document}